\documentclass[conference]{IEEEtran}

\usepackage[english]{babel}
\usepackage{amssymb,textcomp}
\usepackage[usenames]{color}
\usepackage{listings}
\usepackage{paralist}
\usepackage{url}
\usepackage{cite}
\usepackage{xspace}
\usepackage{graphicx}
\usepackage{multirow}
\usepackage{makecell}
\usepackage{fourier} 
\usepackage{array}
\usepackage{array,booktabs}
\usepackage[pdftex,colorlinks=true,pdfstartview=FitV,
 linkcolor=black,citecolor=black,urlcolor=black,bookmarks=false]{hyperref}
\usepackage{needspace}

\usepackage{tikz}
 
\usepackage{ifthen}
\usepackage[normalem]{ulem} 
\usepackage{xcolor}

\newboolean{showedits}
\setboolean{showedits}{true} 
\ifthenelse{\boolean{showedits}}
{
	\newcommand{\del}[1]{\textcolor{red}{\sout{#1}}} 
}{
	\newcommand{\del}[1]{} 
	
}
\usepackage{amssymb}
\newboolean{showcomments}
\setboolean{showcomments}{true}
\newcommand{\id}[1]{$-$Id: scg-llncs.tex 30911 2010-02-05 10:21:47Z oscar $-$}

\ifthenelse{\boolean{showcomments}}
{\newcommand{\nbc}[3]{
 {\colorbox{#3}{\bfseries\sffamily\scriptsize\textcolor{white}{#1}}}
 {\textcolor{#3}{\sf\small$\blacktriangleright$\textit{#2}$\blacktriangleleft$}}}
 }
{\newcommand{\nbc}[3]{}
 \renewcommand{\del}[1]{} 
 }

\newcommand{\ie}{\emph{i.e.},\xspace}
\newcommand{\eg}{\emph{e.g.},\xspace}
\newcommand{\etal}{\emph{et al.}\xspace}
\usepackage{wasysym}
\usepackage{cite}
\usepackage{nth}
\usepackage{scrextend}
\usepackage[ancient,keeplastbox]{flushend}
\usepackage{wrapfig}
\usepackage{subcaption}
\usepackage{colortbl}
\usepackage[all]{nowidow}
\begin{document}


\title{VISON: An Ontology-Based Approach for \\Software Visualization Tool Discoverability}

 \author{
 \IEEEauthorblockN{Leonel Merino\IEEEauthorrefmark{1}, Ekaterina Kozlova\IEEEauthorrefmark{2}\IEEEauthorrefmark{3}, Oscar Nierstrasz\IEEEauthorrefmark{3}, Daniel Weiskopf\IEEEauthorrefmark{1}}
 \IEEEauthorblockA{
 \IEEEauthorrefmark{1}VISUS,
 University of Stuttgart, Germany}
 \IEEEauthorblockA{
 \IEEEauthorrefmark{2}National Research University Higher School of Economics, Russia}
 \IEEEauthorblockA{
 \IEEEauthorrefmark{3}SCG,
 University of Bern, Switzerland}
}

\maketitle
\begin{abstract}
Although many tools have been presented in the research literature of software visualization, there is little evidence of their adoption. To choose a suitable visualization tool, practitioners need to analyze various characteristics of tools such as their supported software concerns and level of maturity. Indeed, some tools can be prototypes for which the lifespan is expected to be short, whereas others can be fairly mature products that are maintained for a longer time. Although such characteristics are often described in papers, we conjecture that practitioners willing to adopt software visualizations require additional support to discover suitable visualization tools. 
In this paper, we elaborate on our efforts to provide such support. To this end, we systematically analyzed research papers in the literature of software visualization and curated a catalog of 70 available tools that employ various visualization techniques to support the analysis of multiple software concerns. We further encapsulate these characteristics in an ontology. \emph{VISON}, our software visualization ontology, captures these semantics as concepts and relationships. We report on early results of usage scenarios that demonstrate how the ontology can support 
\begin{inparaenum}[\itshape (i)\upshape] 
    \item developers to find suitable tools for particular development concerns, and
    \item researchers who propose new software visualization tools to identify a baseline tool for a controlled experiment.
\end{inparaenum}
\end{abstract}

\section{Introduction}
\label{sec:intro}
Complex questions may arise during software development~\cite{Sill06a,Ko07a,Frit10a,Lato10b}. Over the last two decades, many software visualizations have been presented in the research literature and shown to be suitable to address some of these questions~\cite{Meri16a}. 
However, there is still little evidence of where and how software visualizations are being discovered and adopted by practitioners. 
To find a suitable tool, practitioners need to examine aspects such as the development tasks supported by the tool, the required execution environment, the level of maturity of the tool, and whether there is a maintenance plan for future improvements and bug fixes. For example, practitioners can be reluctant to adopt some prototypical visualization tools that often have a short lifespan, and more open to adopt tools that belong to long-term projects and are expected to be maintained for a fairly long time. 
\begin{figure}[tbp]
	\centering
	\includegraphics[width=0.98\linewidth]{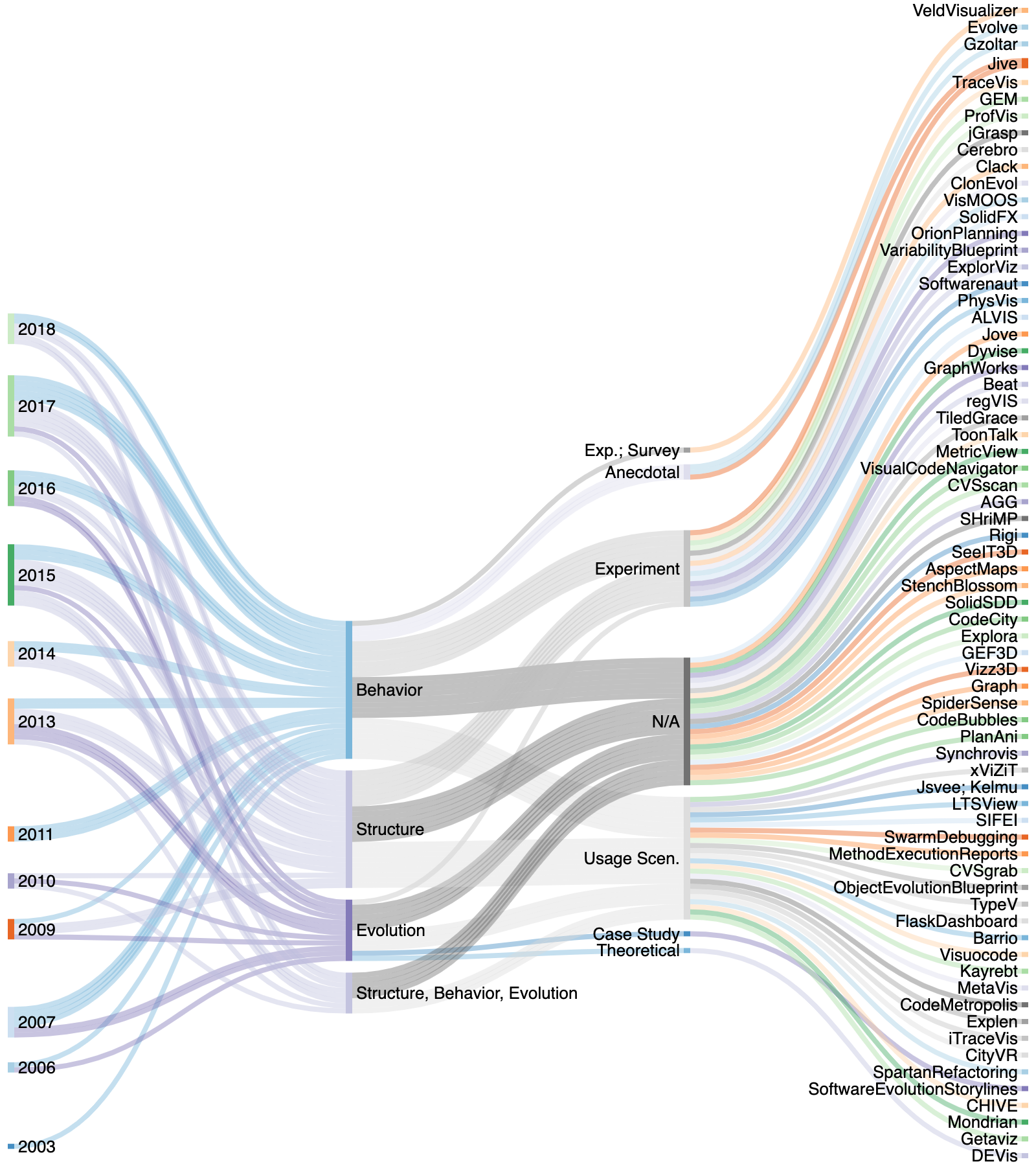}
	\caption{A Sankey diagram that presents our curated catalog of 70 available visualization tools introduced in the literature of software visualization, classified by publication year, software aspects, evaluation strategy, and tool name.}
	\label{fig:tools}
\end{figure} 

\begin{figure*}[tbp]
	\centering
	\includegraphics[width=0.98\linewidth]{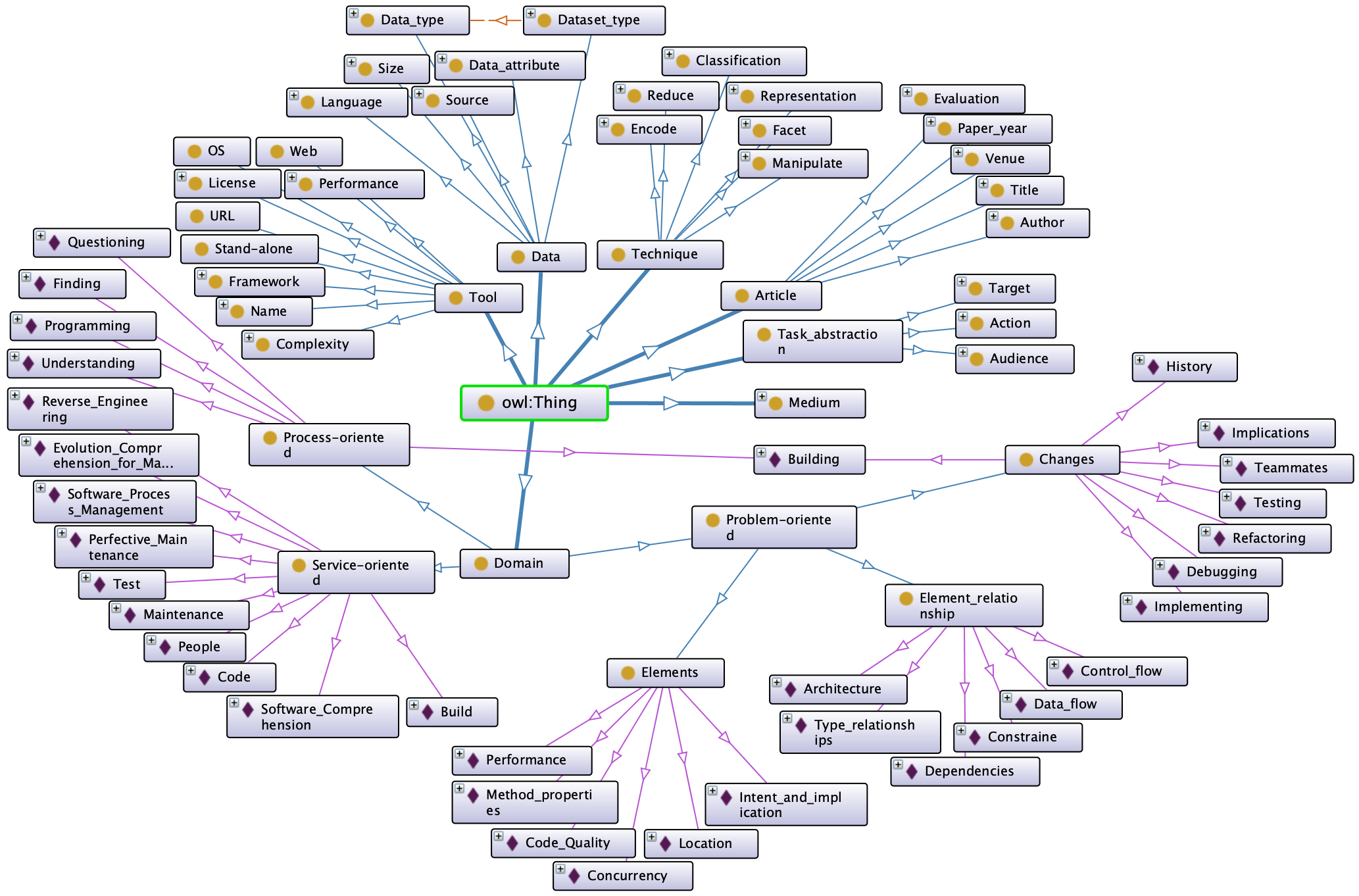}
	\caption{An overview of the concept hierarchy of the \emph{VISON} software visualization ontology using the OntoGraf visualization plug-in. Blue edges denote subclass relationships and violet edges identify instances of a class.}
	\label{fig:protege3}
\end{figure*} 

To address the gap between existing software visualizations and their practical applications, we build on previous studies~\cite{Meri17a,Meri18a} in which we reviewed the literature of software visualization to collect their characteristics. This article is based on results that were reported in the doctoral thesis of the first author~\cite{Meri18b}. We updated the review and curated a catalog of 70 publicly available software visualization tools. For each tool in the catalog, we identify
\begin{inparaenum}[\itshape (i)\upshape] 
    \item the tool's name (\eg Jive),
    \item software aspects (\eg behavior, structure, evolution),
    \item software concerns (\eg execution traces of Java programs), 
    \item last update (\eg 2017),
    \item execution environment (\eg Eclipse plug-in), 
    \item employed visualization techniques (\eg node-link diagram), 
    \item display medium (\eg standard computer screen (SCS), immersive virtual reality (I3D)), and 
    \item evaluations (\eg controlled experiment). 
\end{inparaenum} 
Figure~\ref{fig:tools} shows a Sankey diagram of our catalog of visualization tools introduced in the literature of software visualization. 
The ontology, which enables both textual and visual search methods, then can be used by practitioners to find suitable software visualizations as well as by researchers who can reflect on the software visualization domain. The ontology can also enable higher-level frameworks to support practitioners to search for software visualization tools. 

\emph{Why a software visualization ontology?} Ontologies are formal and explicit descriptions of concepts in a domain~\cite{Grub95a}. Ontologies can help 
\begin{inparaenum}[\itshape (i)\upshape] 
	\item share a common understanding of the structure of information among people or software agents, 
	\item reuse domain knowledge, 
	\item enforce domain assumptions,
	\item separate domain knowledge from operational knowledge, and 
	\item analyze domain knowledge. 
\end{inparaenum}

Figure~\ref{fig:protege3} shows an overview of \emph{VISON}, our software visualization ontology that encapsulates main characteristics of software visualizations. To the best of our knowledge, VISON is the first ontology of software visualizations. We elaborate on lessons learned from developing the ontology, and early results of usability through usage scenarios.

To populate VISON, we built on a set of selected papers of previous surveys of the software visualization literature~\cite{Meri17a,Meri18a}. Specifically, we scanned each classified design study paper to identify software visualization tools. For each tool that we found, we checked whether the tool is publicly available on the internet. In the end, we curated a catalog of 70 publicly available software visualization tools that we used to populate our ontology.  

The main contribution of this paper is twofold:
\begin{inparaenum}[\itshape (i)\upshape] 
	\item a curated catalog of 70 available software visualization tools and,
	\item a publicly available~\cite{Meri19c} software visualization ontology.
\end{inparaenum}

The remainder of the paper is structured as follows: Section \ref{sec:related} describes related work that focuses on practical applications of software visualizations and catalogs of software visualization tools. Section \ref{sec:background} elaborates on the main concepts of ontologies that are addressed in our study. Section \ref{sec:tools} presents VISON, our software visualization ontology. We first elaborate on a catalog of 70 available software visualization tools, and then we discuss ontology implementation details. Section \ref{sec:conclusion} concludes and presents future work.

\section{Related Work}
\label{sec:related}
We group related work into two main themes. We first discuss research that proposes approaches to practical applications of software visualization tools. Then, we elaborate on studies that present catalogs of software visualization tools.

\subsection{Practical Applications of Software Visualizations} 
We observe that there are only a few studies that have been carried out to fill the gap between existing software visualizations and their practical applications. For instance, Hassaine \etal~\cite{Hass09b} elaborate on an approach for generating visualizations to support software maintenance tasks. Sfayhi and Sahraoui~\cite{Sfay11a} describe how to generate interactive visualizations based on descriptions of code analysis tasks. To this end, developers are required to describe the task using a domain-specific language. Grammel \etal~\cite{Gram10a} investigate how novices construct information visualizations. Based on the analysis of the usage of simple visualizations such as charts and scatter plots, they identify a user's need for information visualization tools. However, we observe that these visualizations provide limited support for the analysis of development concerns. Three other studies~\cite{Gall05a,Pare14a,Shah14a} investigate software development tasks for which visualization tools have been proposed, however, we consider that the tasks in these studies are at a too high-level for developers to find an appropriate visualization to their particular needs. Merino \etal~\cite{Meri16c} introduce a meta-visualization approach of live visualization example objects annotated with the type of development questions that they can help investigate. In the visualization, developers can identify suitable visualization examples by detecting the surrounding keywords in the tag-iconic cloud-based visualization. Instead, we propose the use of an ontology that can encapsulate the semantics of the characteristics of software visualizations. As opposed to the described studies, our ontology-based approach leverages existing software visualization tools by attempting to provide practitioners a means for discovery.    

\subsection{Catalogs of Software Visualization Tools}
Some studies examine software visualization tools, in particular, to create guidelines for designing and evaluating software visualizations. For example, 
Storey~\etal~\cite{Stor05a} examine 12 software visualization tools and propose a framework to evaluate software visualizations based on intent, information, presentation, interaction, and effectiveness.
Sensalire~\etal~\cite{Sens08a,Sens09a} classify the features users require in software visualization tools. To this end, they elaborate on lessons learned from evaluating 20 software visualization tools and identify dimensions that can help design an evaluation and then analyze the results. In our investigation, we do not attempt to provide a comprehensive catalog of software visualization tools, but we seek to provide a means to boost software visualization discoverability.

Some other studies present taxonomies that characterize software visualization tools.
Myers~\cite{Myer90a} classifies software visualization tools based on whether they focus on code, data, or algorithms; and whether they are implemented in a static or dynamic fashion.
Price \etal~\cite{Pric93a} present a taxonomy of software visualization tools based on six dimensions: scope, content, form, method, interaction, and effectiveness.
 Maletic \etal~\cite{Male02a} propose a taxonomy of five dimensions to classify software visualization tools: tasks, audience, target, representation, and medium. Schots \etal~\cite{Scho14a} extend this taxonomy by adding two dimensions: resource requirements of  visualizations, and evidence of their utility. Merino \etal~\cite{Meri17a} add \emph{needs} as a main characteristic of software visualization tools. In their context, ``needs'' refers to the set of questions that are supported by software visualization tools. Although we consider these studies crucial for reflecting on the software visualization domain, we think that practitioners may require a more comprehensive support to identify a suitable tool. In particular, we believe that the semantics of concepts and their relationships are often missing in taxonomies and other classifications. The use of an ontology enforces the analysis of these relationships, which can play an important role in identifying a suitable visualization tools.    

\section{Ontology Design Considerations}
\label{sec:background}
An ontology is a formalization of a model to describe what is essential in a \emph{domain}. That is, the ontology describes the \emph{concepts} in the domain that can define various \emph{properties} and \emph{restrictions}. Hence, an ontology populated with a set of individual \emph{instances} of the concepts is usually referred to as a knowledge base. However, defining what in the domain is modeled as a concept or an instance is subjective. We opted to follow the widely used guidelines proposed by Noy and McGuiness~\cite{Noy01a}. We now elaborate on how we addressed their suggested steps to create our \emph{software visualization ontology}.

\vspace{0.2cm}
\noindent\emph{\textbf{Step 1.} Determine the domain and scope of the ontology.}    
\vspace{0.2cm}

\begin{itemize}
	\setlength{\itemsep}{1pt}
  	\setlength{\parskip}{0pt}
  	\setlength{\parsep}{0pt}
	\item \emph{What is the domain that the ontology will cover?} Software visualizations.
	\item \emph{For what we are going to use the ontology?} To allow 1)~developers to find suitable visualizations for their particular concerns and 2)~researchers to reflect on the software visualization domain.
	\item \emph{For what types of questions the information in the ontology should provide answers?} Questions that identify particular software visualizations that fulfill the restrictions imposed by the context of the developers needs.
	\item \emph{Who will use and maintain the ontology?} Software developers willing to adopt visualizations, and who have used a visualization from the ontology and want to add new supported questions to it. Also, researchers who want to add new data to the ontology for a new or an existing indexed visualization approach.
\end{itemize}

\noindent\emph{\textbf{Step 2.} Consider reusing existing ontologies.} To the best of our knowledge, this is the first ontology of software visualizations.
\vspace{0.2cm}

\noindent\emph{\textbf{Step 3.} Enumerate important terms in the ontology.} We include the characteristics of software visualization and their evaluations, as well as the classifications presented in previous studies~\cite{Meri17a,Meri18a}.
\vspace{0.2cm}

\noindent\emph{\textbf{Step 4.} Define the concepts and the concept hierarchy.} We opt for a bottom-up development process in which we start from instances of proposed software visualizations. For each, we identify the various concepts involved in its context (\eg tasks, media, environments, frameworks, questions, evaluation strategies). We define a hierarchy of concepts following an ``is-a'' relation. When defining the concepts, we avoid creating cycles and validate that sibling concepts (\ie at the same level in the hierarchy) correspond to the same level of generality. 
\vspace{0.2cm}

\noindent\emph{\textbf{Step 5.} Define the properties of concepts.} We characterize the concepts based on their properties. For instance, for the concept \emph{medium} we define the \emph{dimensionality} (\eg 2D/3D) property. Then, when we define particular software visualizations as instances in the ontology, we can specify a medium and its dimensionality. Thus, researchers can use the ontology to investigate, for instance, the correlation between evaluation strategies and visualizations that use visualization techniques of a higher dimensionality displayed on a medium of a lower dimensionality.        
\vspace{0.2cm}

\noindent\emph{\textbf{Step 6.} Define the restrictions of the properties.} We only use restrictions to define disjoint concepts.
\vspace{0.2cm}

\noindent\emph{\textbf{Step 7.} Create instances.} We create instances in the ontology for each proposed software visualization in our data set. Thus, visualization tools are the materialization of a combination of property values of concepts.   

\vspace{1em}

\section{VISON: Software Visualization Ontology}
\label{sec:tools}
Certainly, an empty ontology that describes concepts and relationships but has no instances cannot be useful for practitioners. Therefore, before we describe an implementation of our ontology, we elaborate on the systematic approach that we used to populate it. In the following, we describe the process followed to collect a set of relevant software visualization tools and their characteristics from the research literature.  

\subsection{Software Visualization Tools}
We built on the data sets from the proposed software visualization approaches (presented in previous studies~\cite{Meri17a,Meri18a}). We reviewed the 387 software visualization papers published in the VISSOFT/SOFTVIS conferences. Since the goal of our investigation is to facilitate the discoverability of software visualization tools, we included in our catalog only software visualization tools that are:
\begin{inparaenum}[\itshape (C1)\upshape] 
	\item identified with a name and
	\item publicly available on the internet.
\end{inparaenum}

We scanned each paper to identify a name for the proposed software visualization approach. Then, we looked for a URL where the tool might be available. In most cases (where we did not find a URL in the paper), we searched the Web using the name of the tool (C1). When we did not find a positive result, we added ``visualization'' to the search keywords. When we found an available tool (C2), we checked the last time when the tool was updated. Sometimes, we had to download the tool to look for the date in the files. In the end, we found 70 software visualization tools that fulfill the criteria and that we therefore included in our catalog. 

To characterize a tool we first identified its name and whether it focuses on the structure, behavior, or evolution of software systems~\cite{Dieh07a}. Then, for the tools in each category, we identified the development concern expressed by the visualization. Instead of describing high-level tasks (\eg reverse-engineering), we formulated descriptions with the main keywords of  the concerns (\eg ``reports that summarize methods execution''), which we think can help developers relate their particular context to the one envisioned by a proposed visualization tool. We also classified the tools based on their execution environment, the employed visualization technique, and the medium used to display them. Finally, we reused the data presented in our previous study~\cite{Meri18a} to highlight the maturity of tools that have proven effective to support the target task through evaluations.

\begin{table*}[tbp]
\centering
\caption{A curated catalog of 70 available software visualization tools. Tools are grouped by aspects: Behavior, Structure, Evolution, and E.-S.-B. (their combination).}
\label{tab:tools}
\setlength\tabcolsep{6pt}
\renewcommand{\arraystretch}{0.9}
\begin{tabular}{p{0.2cm}p{3.2cm}p{0.4cm}lp{1.3cm}p{2.3cm}p{0.4cm}p{1.5cm}}
\toprule
\textbf{Asp.} &\textbf{Tool's Name} & \textbf{Year} & \textbf{Software Concern} & \textbf{Environment} & \textbf{Technique} & \textbf{Med.} & \textbf{Evaluation}  \\ \midrule
\multirow{28}{*}{\rotatebox[origin=c]{90}{Behavior}} & \href{https://github.com/danwent/clack-graphical-router}{Clack} & 2018 & Concepts for teaching networks in CS & Java & Node-link & SCS & Anecdotal \\ 
 & \href{https://github.com/ToonTalk}{ToonTalk} & 2018 & Concepts for teaching  children to program & Web & Visual language & SCS & N/A \\ 
 & \href{http://www.mcrl2.org/web/user_manual/tools/release/ltsview.html}{LTSView} & 2017 & Transition systems & Various & 3D node-link & SCS & N/A \\ 
 & \href{http://www.gzoltar.com}{Gzoltar} & 2017 & Fault localization for debugging Java progs. & Java;Ecli. & Icicle; treemap & SCS & Experiment \\ 
 & \href{https://github.com/kuleszdl/SIFEI}{SIFEI} & 2017 & Spreadsheets formulas for testing & Excel & Visual language & SCS & Experiment \\ 
 & \href{https://github.com/SwarmDebugging}{SwarmDebugging} & 2017 & Reuse knowledge of debugging sessions & Eclipse & Node-link & SCS & Usage Scen. \\ 
 & \href{https://github.com/fabian-beck/Method-Execution-Reports}{MethodExecutionReports} & 2017 & Summarization of methods execution & Java & Charts & SCS & Experiment \\ 
 & \href{http://www.cse.buffalo.edu/jive}{Jive} & 2016 & Execution traces of Java programs & Eclipse & Node-L; aug.src. & SCS & Anecdotal \\ 
 & \href{http://spideruci.github.io/cerebro/}{Cerebro} & 2016 & Execution traces for feature identification & Web & Node-link & SCS & Usage Scen. \\ 
 & \href{https://github.com/Aalto-LeTech/jsvee}{Jsvee; Kelmu} & 2016 & Concepts for teaching programming in CS & Web & Aug. source code & SCS & Usage Scen. \\ 
 & \href{https://github.com/lgeorget/Kayrebt-Dumper}{Kayrebt} & 2015 & Control and data flow of the Linux kernel & Linux & Node-link & SCS & Usage Scen. \\ 
 & \href{http://homepages.ecs.vuw.ac.nz/~mwh/}{TiledGrace} & 2015 & Programming in the Grace language & Web & Visual language & SCS & Experiment \\ 
 & \href{http://www.jgrasp.org/}{jGrasp} & 2015 & Concepts for teaching programming in CS & Various & Aug. source code & SCS & Exp.; Survey \\ 
 & \href{https://xvizit.wordpress.com/portfolio/metrics-based-spreadsheet-visualization/}{xViZiT} & 2015 & Spreadsheets formulas for testing & Java & Aug. source code & SCS & Usage Scen. \\ 
 & \href{https://github.com/pj/beat}{Beat} & 2014 & Execution traces of Java concurrent prog. & Eclipse & Aug. source code & SCS & N/A \\ 
 & \href{http://www.sts.tu-harburg.de/projects/regvis/regvis.html}{regVIS} & 2014 & Assembler control-flow of regular expr. & Windows & Visual language & SCS & Experiment \\ 
 & \href{https://github.com/danmedani/GraphWorks}{GraphWorks} & 2013 & Concepts for teaching graph theory in CS & Java & Anim. node-link & SCS & N/A \\ 
 & \href{http://kieker-monitoring.net/download/synchrovis/}{Synchrovis} & 2013 & Execution traces of Java concurrent prog. & Java & City & SCS & Usage Scen. \\ 
 & \href{http://www.cs.uef.fi/~saja/var_roles/planani/index.html}{PlanAni} & 2011 & Concepts for teaching programming in CS & Various & Aug. source code & SCS & Experiment \\ 
 & \href{http://ftaiani.ouvaton.org/7-software/profvis.html}{ProfVis} & 2011 & Execution traces of Java programs & Java & Node-link & SCS & Experiment \\ 
 & \href{http://formalverification.cs.utah.edu/ISP-Eclipse/}{GEM} & 2011 & Dynamic verification of MPI programs & Eclipse & Aug. source code & SCS & N/A \\ 
 & \href{ftp://ftp.cs.brown.edu/u/spr/dyvise.tar.gz}{Dyvise} & 2009 & Java heap to detect memory problems & Java & Icicle & SCS & Anecdotal \\ 
 & \href{http://cs.brown.edu/~spr/research/vizjive.html}{Jive} & 2007 & Execution traces of Java programs & Java & Charts & SCS & Usage Scen. \\ 
 & \href{http://cs.brown.edu/~spr/}{Jove} & 2007 & Execution traces of Java programs & Java & Charts & SCS & N/A \\ 
 & \href{http://cs.brown.edu/~spr/}{VeldVisualizer} & 2007 & Execution traces of Java programs & Java & Pixel & SCS & N/A \\ 
 & \href{http://www.win.tue.nl/~wstahw/projects/finished/PieterDeelen/index.html}{TraceVis} & 2007 & Execution traces based on call graphs & Java & Node-link & SCS & Usage Scen. \\ 
 & \href{http://eecs.wsu.edu/~veupl/soft/index.htm}{ALVIS} & 2006 & Concepts for teaching programming in CS & Windows & Visual language & SCS & N/A \\
 & \href{http://www.sable.mcgill.ca/evolve/}{Evolve} & 2003 & Execution traces of Java programs & Java & Pixel & SCS & Usage Scen. \\ [2ex] 
\multirow{22}{*}{\rotatebox[origin=c]{90}{Structure}}  & \href{https://bitbucket.org/physviz/physviz}{PhysVis} & 2018 & Software quality based on metric analysis & VisualStudio & 3D node-link & I3D & Usage Scen. \\ 
 & \href{https://marketplace.eclipse.org/content/spartan-refactoring-0}{SpartanRefactoring} & 2018 & Automatic code refactoring for readability  & Eclipse & Aug. source code & SCS & N/A \\ 
 & \href{https://github.com/mircealungu/Softwarenaut}{Softwarenaut} & 2017 & Architecture and dependency analysis & VisualWorks & Node-L.; treemap & SCS & N/A \\ 
 & \href{https://github.com/geryxyz/CodeMetropolis}{CodeMetropolis} & 2017 & Software quality based on metric analysis & Java & City & SCS & N/A \\ 
 & \href{https://github.com/arnobl/kompren}{Explen} & 2017 & Slice-based techs. for large metamodels & Eclipse & UML & SCS & N/A \\ 
 & \href{https://github.com/SERESLab/iTrace-Archive}{iTraceVis} & 2017 & Eye movement data of code reading & Eclipse & Heatmap & SCS & Experiment \\ 
 & \href{http://scg.unibe.ch/research/cityvr}{CityVR} & 2017 & Architecture based on metrics in OOP & Pharo; U. & City & I3D & Experiment \\ 
 & \href{https://www.explorviz.net}{ExplorViz} & 2016 & Architecture based on metric analysis & Web & City & S/I & Experiment \\ 
 & \href{http://smalltalkhub.com/#!/~merino/MetaVisualization}{MetaVis} & 2016 & Annotated visualization example objects & Pharo & Node-L.; tag cloud & SCS & Usage Scen. \\ 
 & \href{http://smalltalkhub.com/#!/~RichardWettel/CodeCity}{CodeCity} & 2015 & Software quality based on code smells & Pharo & City & SCS & Usage Scen. \\ 
 & \href{http://scg.unibe.ch/research/explora}{Explora} & 2015 & Software quality based on metric analysis & Pharo & Polymetric views & SCS & Usage Scen. \\ 
 & \href{http://smalltalkhub.com/#!/~GustavoSantos/OrionChangesBrowser}{OrionPlanning} & 2015 & Arch. modularization and consistency & Pharo & Node-link & SCS & Usage Scen. \\ 
 & \href{http://smalltalkhub.com/#!/~abergel/Familiar}{VariabilityBlueprint} & 2015 & Decomposition of models in FOP & Pharo & Polymetric views & SCS & Usage Scen. \\ 
 & \href{https://github.com/DeveloperLiberationFront/refactoring-tools/tree/master/installables/update_sites/stench_blossom}{StenchBlossom} & 2014 & Software quality based on code smells & Eclipse & Aug. source code & SCS & Experiment \\ 
 & \href{https://www.visuocode.com}{Visuocode} & 2014 & Navigation and composition of  systems & Mac & Aug. source code & SCS & N/A \\ 
 & \href{http://www.solidsourceit.com/index.html}{SolidSDD} & 2014 & Software quality based on code clones & Windows & HEB & SCS & Usage Scen. \\
 & \href{http://www.solidsourceit.com/products/SolidFX-static-code-analysis.html}{SolidFX} & 2013 & Architecture, metric and dependencies & Windows & HEB; pixel & SCS & Experiment \\ 
 & \href{https://github.com/davidmr/seeit3d}{SeeIT3D} & 2013 & Software architecture of Java systems & Eclipse & City & SCS & Experiment \\ 
 & \href{https://pleiad.cl/research/software/aspectmaps#}{AspectMaps} & 2013 & Architecture of aspect-oriented programs & Pharo & Iconic; pixel & SCS & Experiment \\ 
 & \href{http://ftparmy.com/123154-vismoos.html}{VisMOOS} & 2010 & Software architecture of Java systems & Eclipse & Node-link & SCS & N/A \\ 
 & \href{http://www.rigi.cs.uvic.ca/Pages/download.html}{Rigi} & 2009 & Architecture and dependency analysis & Various & Node-link & SCS & N/A \\ 
 & \href{https://code.google.com/archive/p/barrio/}{Barrio} & 2009 & Architecture and dependency analysis & Eclipse & Node-link & SCS & Usage Scen. \\[2ex] 
 \multirow{12}{*}{\rotatebox[origin=c]{90}{Evolution}}   &\href{https://pypi.org/project/flask-monitoring-dashboard/1.8/}{FlaskDashboard} & 2017 & Flask Python Web services performance & Python & Charts; heatmap & SCS & Usage Scen. \\ 
 &\href{http://smalltalkhub.com/#!/~abergel/ObjectEvolutionBlueprint}{ObjectEvolutionBlueprint} & 2016 & Object mutations & Pharo & Charts & SCS & Experiment \\ 
 &\href{https://github.com/mdfeist/TypeV}{TypeV} & 2016 & Abstract syntax trees of a system's project & Web & Charts & SCS & Usage Scen. \\ 
 & \href{https://sourceforge.net/projects/chiselgroup/}{SHriMP} & 2015 & Hierarchical structures in OOP & Eclipse & Node-link & SCS & N/A \\ 
 &\href{http://www.user.tu-berlin.de/o.runge/AGG/WWW/down_V205/index.html}{AGG} & 2013 & Hierarchical structures in OOP & Java & Node-link & SCS & N/A \\ 
 &\href{https://sites.google.com/site/junjizhi/devis_tool}{DEVis} & 2013 & Technical documents & Eclipse & Spiral & SCS & Theoretical \\ 
 &\href{http://www.cs.rug.nl/svcg/SoftVis/ClonEvol}{ClonEvol} & 2013 & Software quality based on code clones & Windows & HEB & SCS & Usage Scen. \\ 
 &\href{https://code.google.com/archive/p/evolines/downloads}{SoftwareEvolutionStorylines} & 2010 & Developers interactions in projects & Processing & StoryLines; charts & SCS & N/A\\
 &\href{http://www.cs.rug.nl/svcg/SoftVis/VCN}{CVSgrab} & 2009 & Interactions during debugging & Windows & Pixel & SCS & N/A \\ 
 &\href{http://www.win.tue.nl/vis1/home/lvoinea/VCN.html}{VisualCodeNavigator} & 2007 & Source code changes & Windows & Aug. src.; pixel & SCS & Usage Scen. \\ 
 &\href{http://www.win.tue.nl/vis1/home/lvoinea/VCN.html}{CVSscan} & 2007 & Source code changes & Windows & Pixel & SCS & Case Study \\ 
 &\href{http://www.win.tue.nl/san/projects/empanada/metricview/}{MetricView} & 2006 & Hierarchical structures and metrics in OOP & Windows & 3D UML  & SCS & N/A  \\[2ex] 
\multirow{8}{*}{\rotatebox[origin=c]{90}{E.-S.-B.}} & \href{http://agilevisualization.com/AgileVisualization/Mondrian/0202-Mondrian.html}{Mondrian} & 2018 & Execution traces of feature dependencies & Pharo & Polymetric views & SCS & N/A \\
 & \href{https://github.com/getaviz/Getaviz}{Getaviz} & 2018 & Developing and evaluating software vis. & Web & City & S/I & N/A \\ 
 & \href{http://cs.brown.edu/~spr/codebubbles/}{CodeBubbles} & 2018 & Debugging within CodeBubbles & Ecli.; VS & Visual language & SCS & N/A \\ 
 & \href{http://sourceforge.net/projects/chive/}{CHIVE} & 2015 & Feature location (reconnaissance) & Eclipse & 3D node-link & SCS & N/A \\ 
 & \href{http://smalltalkhub.com/#!/~abergel/GraphViewer}{Graph} & 2015 & Code dependencies & Pharo & Node-link & SCS & Usage Scen. \\ 
 & \href{https://github.com/spideruci/sense-vis}{SpiderSense} & 2015 & Execution traces of Java programs & Web & Pixel; treemap & SCS & Usage Scen. \\ 
 & \href{http://vizz3d.sourceforge.net}{Vizz3D} & 2013 & Software architecture and quality & Java & 3D node-link & SCS & N/A \\ 
 & \href{https://www.eclipse.org/gef3d/}{GEF3D} & 2010 & Execution traces of Java programs & Eclipse & 3D UML & SCS & Usage Scen. \\ 
 \bottomrule
\end{tabular}
\end{table*}

Table~\ref{tab:tools} presents our curated catalog of 70 available software visualization tools classified by the software's  \emph{date} of last update, \emph{environment} required to execute, employed visualization \emph{technique}, \emph{medium}, and evidence of the visualization's effectiveness through \emph{evaluations}. Each tool's name is linked to a URL that contains instructions for downloading and installation. Visualization tools are classified into software \emph{aspects}: behavior, evolution, and structure.

\subsubsection{Behavior} 
Several visualization tools support teaching various subjects in computer science. \emph{ToonTalk}~\cite{Kahn06a} comes with a visual language (similar to Scratch~\cite{Malo04a}) that is to be used on the Web. The tool targets children as an audience. We are not aware of any evaluation of ToonTalk. However, the tool has been maintained over the last twelve years, which shows evidence of maturity. Similarly, \emph{Tiled Grace}~\cite{Home13a} offers a visual representation alternative to the textual mode when programming in the Grace language. Another mature tool is \emph{Clack}~\cite{Wend06a}, which helps students of network courses understand the behavior of routers. \emph{GraphWorks}~\cite{Meda10a} focuses on supporting students of graph theory, although it has not been maintained in the last few years.   

\begin{figure}[tbp]
	\centering
	\includegraphics[width=0.98\linewidth, height=6cm]{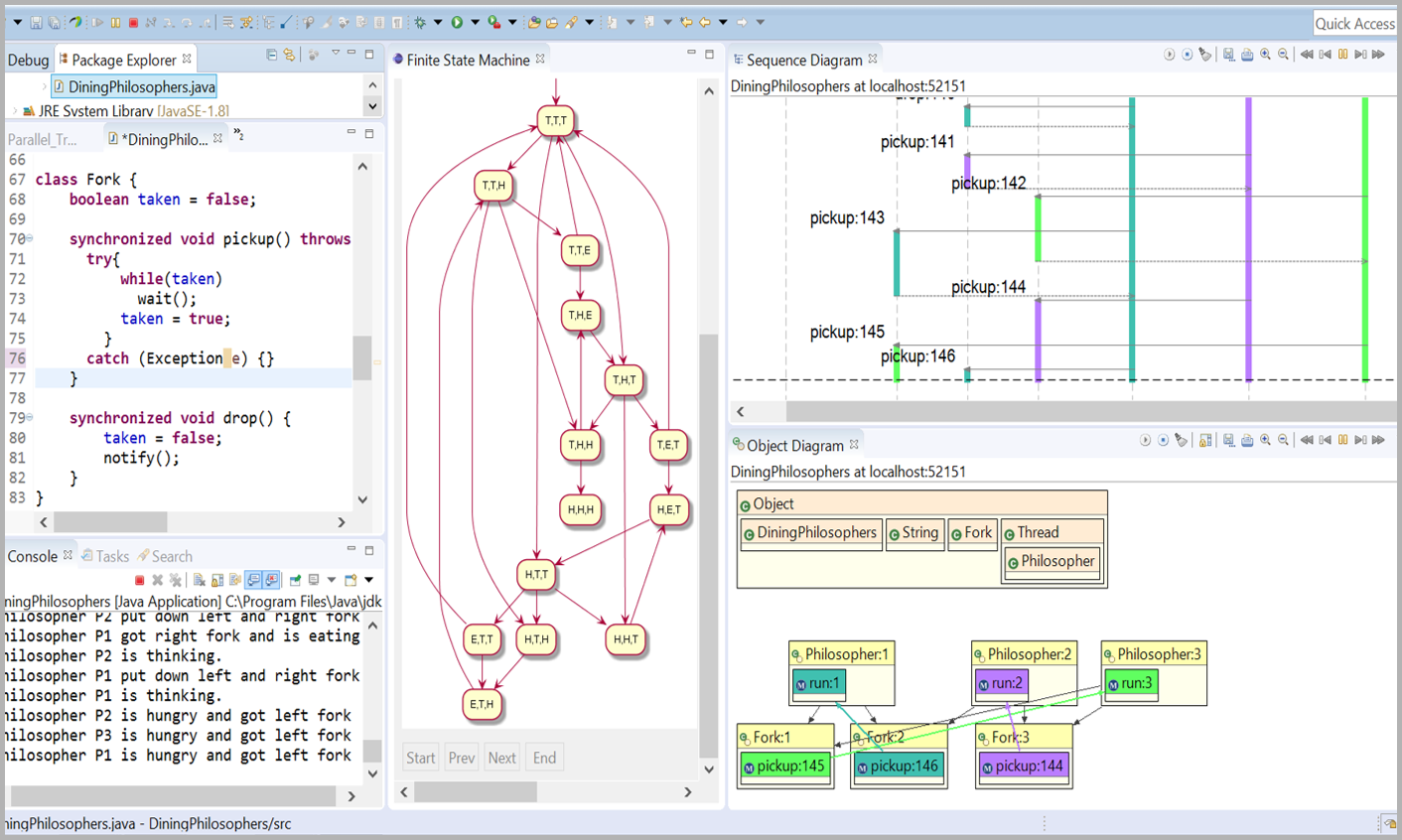}
	\caption{The \emph{Jive} visualization tool to support the analysis of behavior of concurrent Java applications. Figure taken from the Web~\cite{JIVE}, and reused with permission \textcopyright\space~2005 Jayaraman.}
	\label{fig:jive}
\end{figure}

Some other tools are available to deal with understanding the execution of programs for testing. The Eclipse plug-in \emph{Jive}~\cite{Gest05a} (shown in Figure~\ref{fig:jive}) stands out since it has been maintained for the last eleven years, which is congruent with anecdotal evidence of its adoption. Even though all of these tools are available, almost none of them have been maintained lately. Amongst them, \emph{ProfVis}~\cite{Lin10a} is the only one that has proven effective in an experiment. A few others---\emph{Jove}~\cite{Reis05a} and \emph{Veld Visualizer}~\cite{Reis06z}---have been presented only through usage scenarios. Other tools that, to our knowledge, have not been evaluated are \emph{Jive}~\cite{Reis03a}, \emph{TraceVis}~\cite{Deel07a}, and \emph{Evolve}~\cite{Wang03a}.  Two tools---\emph{Beat}~\cite{John10a} and \emph{Synchrovis}~\cite{Wall13a}---target the analysis of the behavior of concurrent Java programs, whereas the tool \emph{Cerebro}~\cite{Pale15a} can be used to identify software features from the runtime data.

\begin{figure}[tbp]
	\centering
	\includegraphics[width=0.98\linewidth]{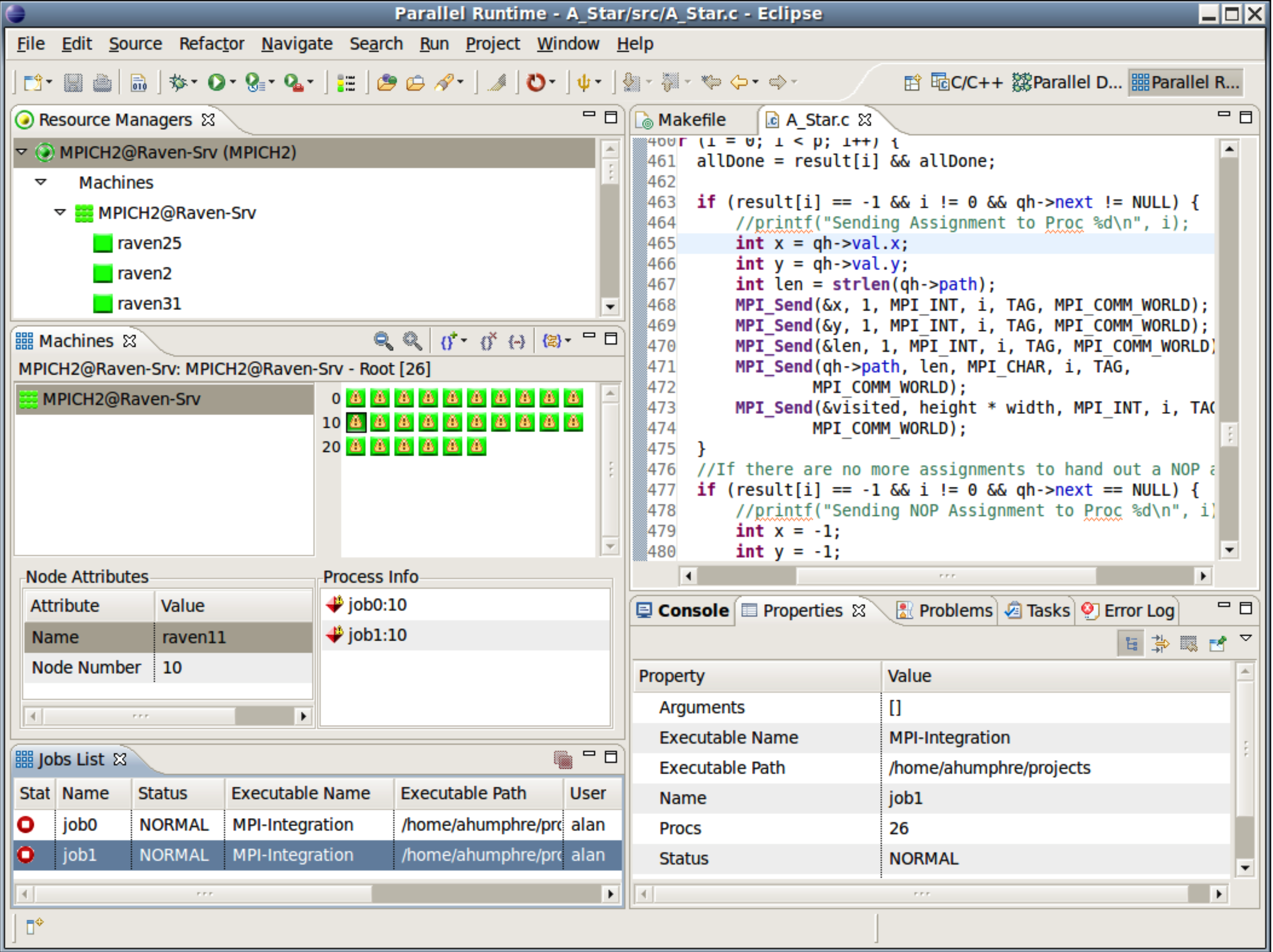}
	\caption{The \emph{GEM} graphical explorer of MPI programs. \textcopyright\space~2010 IEEE. Reprinted, with permission, from Humprey \etal~\cite{Hump10a}.}
	\label{fig:gem}
\end{figure}

Three visualization tools support debugging tasks based on the visualization of program behavior. \emph{Dyvise}~\cite{Reis09z} supports the detection of memory problems through the visualization of the Java heap. \emph{GEM}~\cite{Hump10a} (shown in Figure~\ref{fig:gem}) is a graphical explorer of MPI programs. \emph{Gzoltar}~\cite{Gouv13a} has shown evidence of effectiveness through an experiment. More recently, \emph{SwarmDebugging}~\cite{Petr15a} is an Eclipse plug-in that aims to reuse the knowledge of previous debugging sessions to recommend locations in the code to define breakpoints. 

Four visualization tools support various aspects of teaching computer science courses. \emph{PlanAni}~\cite{Saja03a} is a program animation system based on the concept of the roles of variables for teaching programming. \emph{ALVIS}~\cite{Hund06a} enables algorithm visualization to learning programming. \emph{jGrasp}~\cite{Cros07a} is an integrated development environment with visualizations for teaching Java. \emph{Jsvee} and \emph{Kelmu}~\cite{Sirk18a} are visualization libraries to help instructors teach aspects of the runtime of a program.

Other visualization tools deal with various particular concerns. \emph{LTSView}~\cite{Ploe08a} is the oldest one, and it is still being maintained. It supports the visualization of transition systems that model the behavior of a software. \emph{SIFEI}~\cite{Kule14a} and \emph{xViZiT}~\cite{Hodn15a} focus on the visualization of spreadsheets, whereas \emph{regVis}~\cite{Topr14a} deals with the visualization of assembler control-flow based on regular expressions. \emph{Method Execution Reports}~\cite{Beck17z} embed word-size graphics in reports of method executions. \emph{Kayrebt}~\cite{Geor15a} provides support for activity diagram extraction and a visualization toolset designed for the Linux codebase.

All the twenty-eight listed tools that focus on the behavior of software systems are displayed on the standard computer screen.

\subsubsection{Structure} 

Various other visualization tools focus on particular concerns. \emph{MetaVis}~\cite{Meri16c} can be used to visualize annotated software visualization example objects. \emph{OrionPlanning}~\cite{Sant15a} includes visualization for modularization and consistency of software projects. \emph{Explen}~\cite{Blou14a} supports the visualization of large metamodels. \emph{iTraceVis}~\cite{Clar17a} has shown evidence of being effective to investigate how developers read code through the visualization of their eye gazes. \emph{Spartan Refactoring} allows automatic code refactoring in the editor. \emph{Visuocode}~\cite{Brad13a} supports the navigation and composition of software systems.  

\begin{figure}[tbp]
	\centering
	\includegraphics[width=1\linewidth]{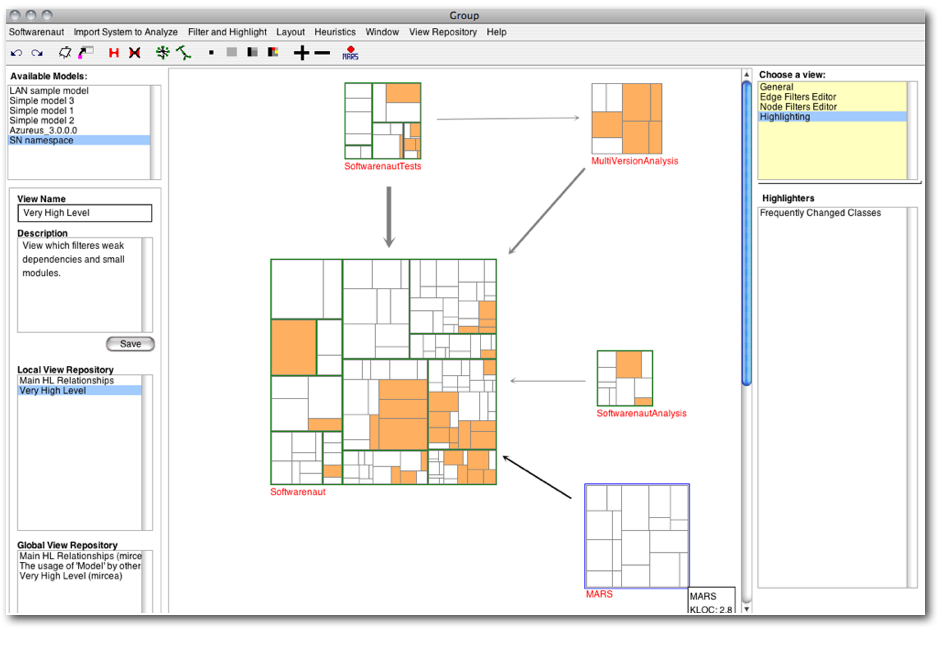}
	\caption{The \emph{Softwarenaut} tool for visualization of hierarchical structures to support architecture tasks. Figure taken from the Web~\cite{Softwarenaut}, and reused with permission~\textcopyright\space~2006 Lungu.}
	\label{fig:softwarenaut}
\end{figure}

Some visualization tools are available for supporting architecture tasks such as \emph{SeeIT3D}~\cite{Shar13a} and \emph{VisMOOS}~\cite{Fron06a}. \emph{SolidFX}~\cite{Tele08a}, \emph{Softwarenaut}~\cite{Lung06a} (shown in Figure~\ref{fig:softwarenaut}), \emph{Rigi}~\cite{Kien07a}, and \emph{Barrio}~\cite{Diet08a} are suitable for the analysis of structures and dependencies in object-oriented software systems, \emph{AspectMaps}~\cite{Fabr13a} supports aspect-oriented programs, and \emph{Variability blueprint}~\cite{Urli15a} does so for feature-oriented programs. 

\begin{figure}[tbp]
	\centering
	\includegraphics[width=0.98\linewidth]{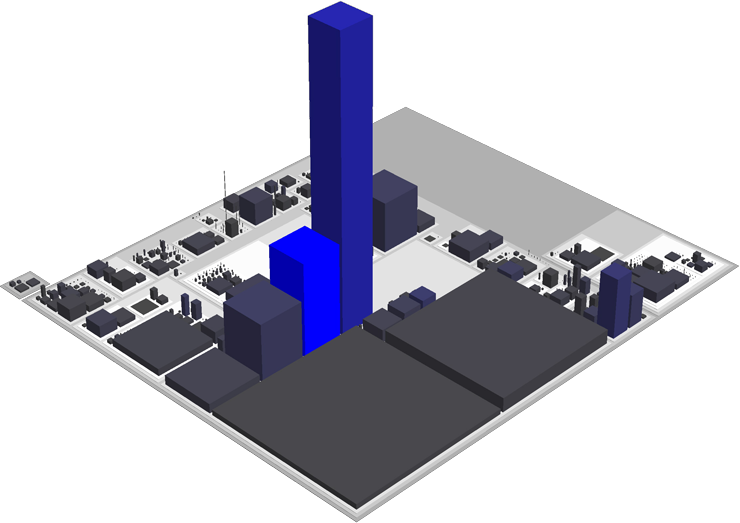}
	\caption{The \emph{CodeCity} tool, which visualizes the structure of software systems to support the analysis of code smells. Figure taken from the Web~\cite{CodeCity}, and reused with permission~\textcopyright\space~Wettel.}
	\label{fig:codecity}
\end{figure}

Three tools support the visualization of software systems quality based on the analysis of code smells. \emph{CodeCity}~\cite{Wett07a} (shown in Figure~\ref{fig:codecity}) and \emph{CodeMetropolis}~\cite{Balo13a} visualize software metrics based on the city metaphor. \emph{StenchBlossom}~\cite{Murp10a} augments the Eclipse source code editor with ambient visualizations. \emph{SolidSDD}~\cite{Voin14a} supports visual clone analysis. \emph{Explora}~\cite{Meri15b} is a visualization tool for software quality based on the analysis of metrics.

Twenty of the listed tools that focus on the structure of software systems are displayed using the standard computer screen. Only three use immersive virtual reality: \emph{PhysVis}~\cite{Scar15a}, in which users visualize software metrics shown as a physical particle system, \emph{ExplorViz}~\cite{Fitt13a}, in which developers obtain an overview of the architecture of a system represented as a city, and \emph{CityVR}~\cite{Meri17c}, which adds interactions and visualization of software metrics and smells. 

\subsubsection{Evolution} 

\begin{figure}[tbp]
	\centering
	\includegraphics[width=0.98\linewidth]{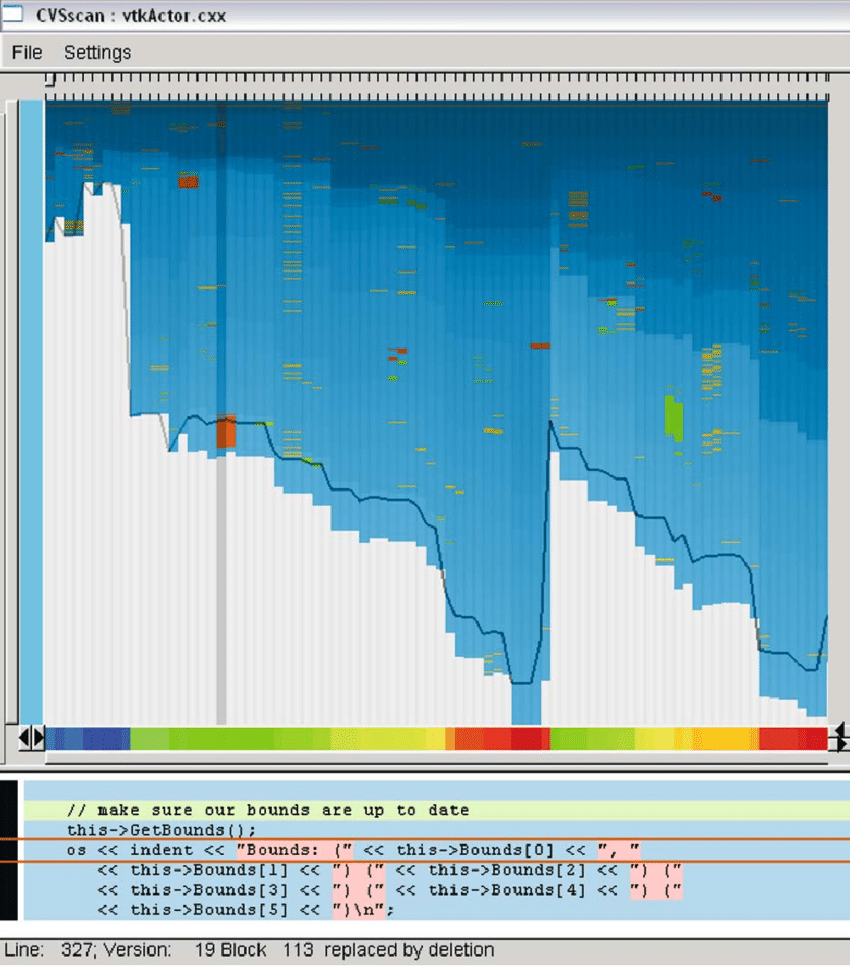}
	\caption{The \emph{CVSscan} visualization tool to support the analysis of evolution for software maintenance. Figure taken from the Web~\cite{CVSScan}, and reused with permission~\textcopyright\space~2005 Telea.}
	\label{fig:cvsscan}
\end{figure}

\begin{figure}[tbp]
	\centering
	\includegraphics[width=0.98\linewidth]{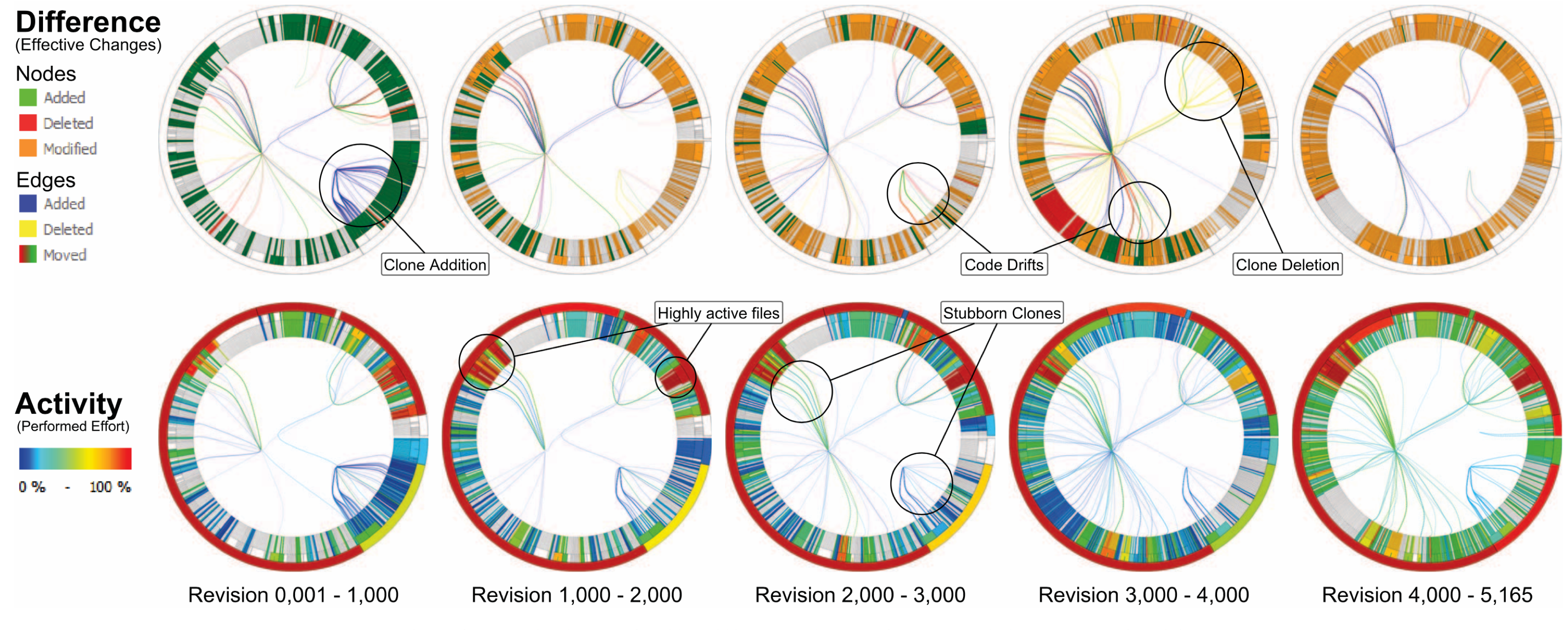}
	\caption{The \emph{ClonEvol} visualization tool, which helps developers analyze the evolution of code clones. \textcopyright\space~2013 IEEE. Reprinted, with permission, from Hanjali\'{c}~\cite{Hanj13a}.}
	\label{fig:clonevol}
\end{figure}

A few tools support the visualization of the evolution of hierarchical structures in object-oriented programs such as \emph{AGG}~\cite{Juck06a}. \emph{SHriMP}~\cite{Lint03z} is the oldest one and has been maintained for twelve years. \emph{MetricView}~\cite{Term05z} presents a UML class diagram in 3D that is augmented with software metrics. Others deal with various concerns. \emph{CVSgrab}~\cite{Voin06z} supports the visualization of the evolution of interactions of developers during debugging, whereas \emph{Visual Code Navigator}~\cite{Depa06a} and \emph{CVSscan}~\cite{Voin05a} (shown in Figure~\ref{fig:cvsscan}) focus on source code changes. \emph{DEVis}~\cite{Zhi13a} is used to visualize the evolution of technical documents. \emph{Object Evolution Blueprint}~\cite{Schu16a} deals with the evolution of object mutations. \emph{Flask dashboard}~\cite{Voge17a} supports the visualization of the performance evolution over versions of Web services implemented using the Flask framework for Python. \emph{TypeV}~\cite{Feis16a} allows one to analyze the evolution of a system through the visualization of abstract syntax trees. \emph{ClonEvol}~\cite{Hanj13a} (shown in Figure~\ref{fig:clonevol}) visualizes the evolution of code clones to improve the quality of systems. \emph{Software Evolution Storylines}~\cite{Ogaw10a} supports the visualization of the interactions between developers during software project evolution.   

All the twelve listed tools that focus on the evolution of software systems are displayed on the standard computer screen.

\subsubsection{Behavior/Evolution/Structure} 

Eight approaches correspond to frameworks that can be used to visualize multiple aspects of software systems. Four of them correspond to active projects introduced several years ago. \emph{Mondrian}~\cite{Meye06a} is an engine for rapid lightweight visualization, which is currently supported in the Roassal engine~\cite{Berg16a}. 
\emph{CodeBubbles}~\cite{Reis14a} is an environment that encapsulates code snippets into bubbles that can be reused through composition. \emph{Vizz3D}~\cite{Pana05a} is a framework for online configuration of 3D information visualizations that was originally available for Eclipse, and later made available for Visual Studio. \emph{CHIVE}~\cite{Clea05a} is a framework for developing, in particular, 3D software visualizations.

\begin{figure}[tbp]
	\centering
	\includegraphics[width=0.69\linewidth]{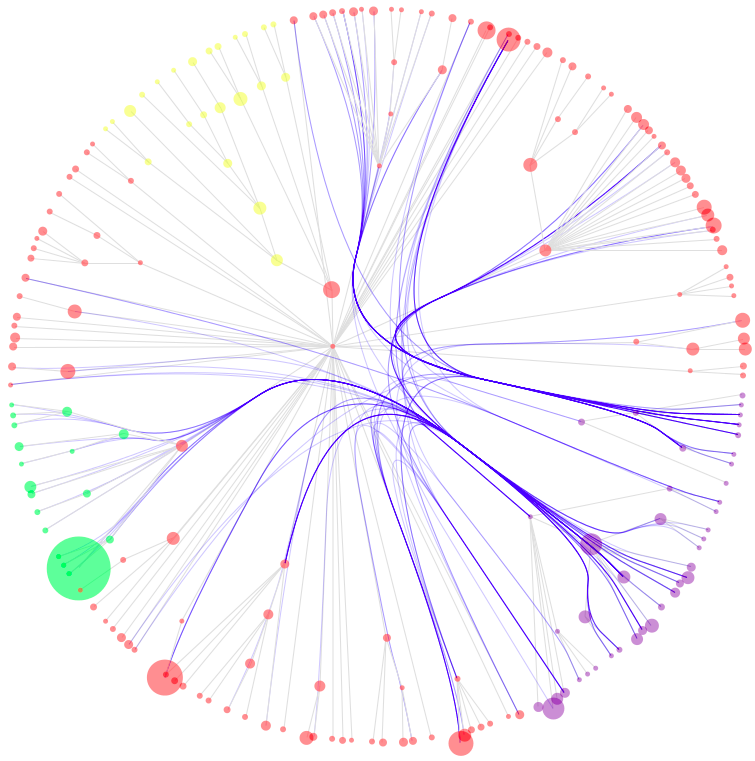}
	\caption{The \emph{Graph} domain-specific language for agile prototyping of visualization of graph structures. Figure taken from the Web~\cite{GRAPH}, and reused with permission~\textcopyright\space~2014 Bergel.}
	\label{fig:graph}
\end{figure}

\emph{GEF3D}~\cite{Pilg08a} is a framework for developing 2D/2.5D/3D graphical editors. \emph{Graph}~\cite{Berg14a} is a domain-specific language for visualizing software dependencies as a graph (shown in Figure~\ref{fig:graph}). \emph{Getaviz}~\cite{Ba17a} and \emph{SpiderSense}~\cite{Redd15a} enable the design, implementation, and evaluation of software visualizations.

The framework \emph{Getaviz} supports visualizations displayed in immersive virtual reality, whereas the seven other frameworks are limited to the standard computer screen.

\begin{figure}[tbp]
	\centering
	\includegraphics[width=\linewidth]{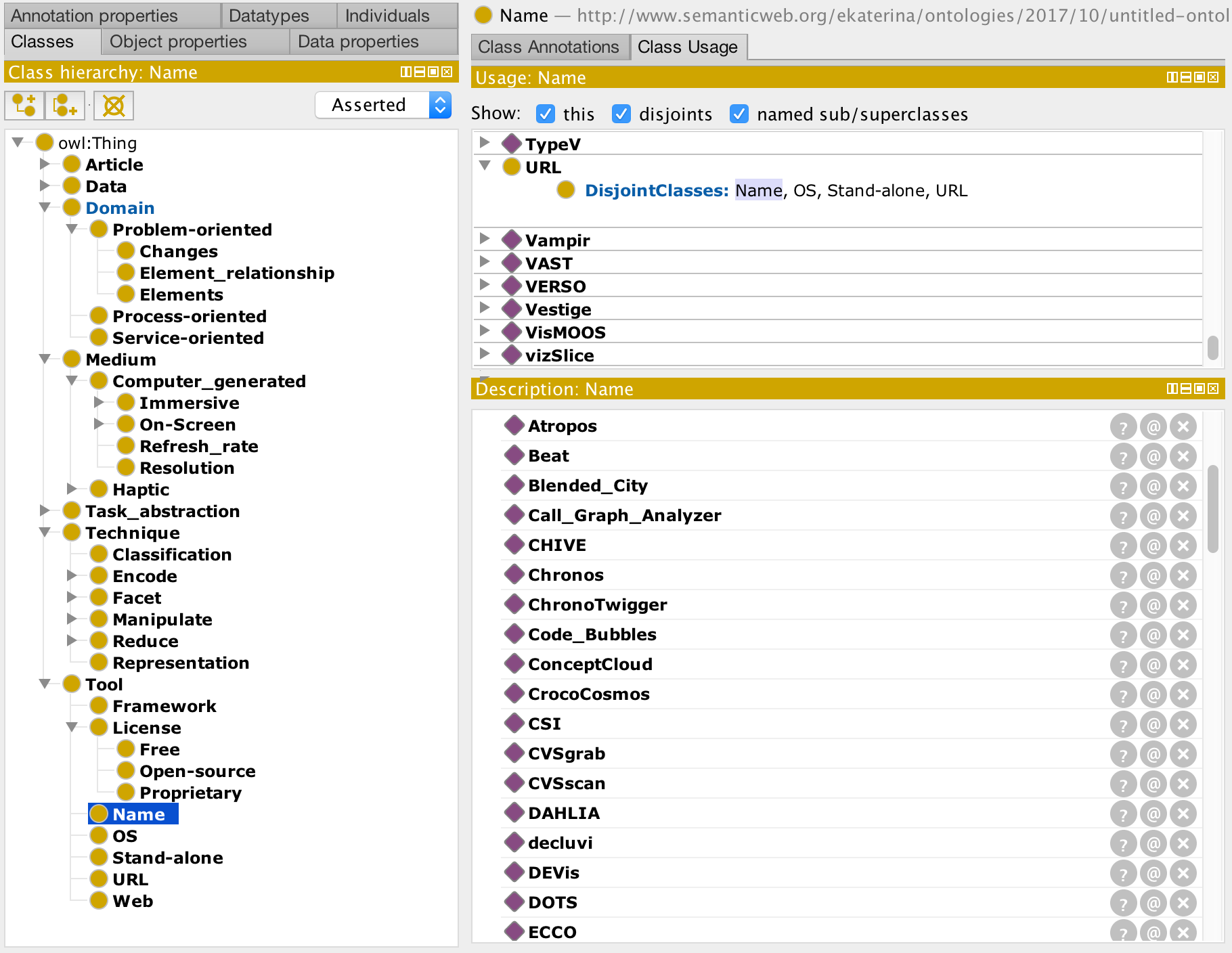}
	\caption{The \emph{classes} view in Prot\'eg\'e showing the hierarchy of concepts. We selected the name of the tools, which are listed in the right pane.}
	\label{fig:protege4}
\end{figure} 


The characterization presented in Table~\ref{tab:tools} contains only part of the content of our data set described in previous publications~\cite{Meri17a,Meri18a}. Various other characteristics of software visualizations can help developers willing to adopt visualization to find a suitable approach. We believe that an ontology can be suitable to implement such richer model.

\subsection{Implementation Details}
\label{sec:vison}
We implement our ontology using \emph{Prot\'eg\'e}~\cite{Muse15a}, a popular, free, and open-source framework for the design and use of ontologies. In it, we define the concepts (in the tool called \emph{classes}), properties, restrictions, and instances. 
Figure~\ref{fig:protege4} shows the \emph{classes} view in Prot\'eg\'e with a detail of the hierarchy of concepts. As the concept, we selected the name of the tools, which are listed in the right pane.

Figure~\ref{fig:protege3} shows an overview of our implementation of the concepts hierarchy using the \emph{OntoGraf} visualization plug-in included in Prot\'eg\'e.

\begin{table}[tbp]
\centering
\caption{Metrics of the software visualization ontology.}
\label{tab:onto:metrics}
\setlength\tabcolsep{4pt}
\begin{tabular}{llr} \toprule 
\textbf{Property} & \textbf{Metric} & \textbf{Value}\\ \midrule 
\multirow{5}{*}{Metrics} & Axiom & 3,290 \\
 & Logical axiom count & 2,428 \\
 & Declaration axiom count & 862 \\
 & Class count & 150 \\
 & Individual property count & 20 \\ \midrule
\multirow{2}{*}{Class axioms} & SubClassOf & 143 \\
 & DisjointClasses & 32 \\  \midrule
\multirow{3}{*}{Object property axioms} & SubObjectPropertyOf & 1 \\
 & ObjectPropertyDomain & 2 \\
 & ObjectPropertyRange & 3 \\  \midrule
\multirow{3}{*}{Individual axioms} & ClassAssertion & 696 \\
 & ObjectPropertyAssertion & 1,547 \\
 & NegativeObjectPropertyAssertion & 4 \\ \bottomrule
\end{tabular}
\end{table}

We present the list of metrics available in the \emph{Ontology metrics} view of Prot\'eg\'e in Table~\ref{tab:onto:metrics}. Although we are aware that many more individuals and relationships must be added to the ontology to increase its usability, we observe that our current implementation is not small according to a survey of ontology metrics~\cite{Sici12a} that reported that ontologies on average contain 36.11 classes (standard deviation of 78.53) and 28.13 instances (standard deviation of 97.59). 

\section{Usage Scenarios}
We now demonstrate the ontology through two usage scenarios.

\vspace{0.2cm}
\noindent\emph{\textbf{Scenario 1.} Find suitable visualization tools that support the analysis of performance issues at runtime.}

\begin{figure}[tbp]
	\centering
	\includegraphics[width=\linewidth]{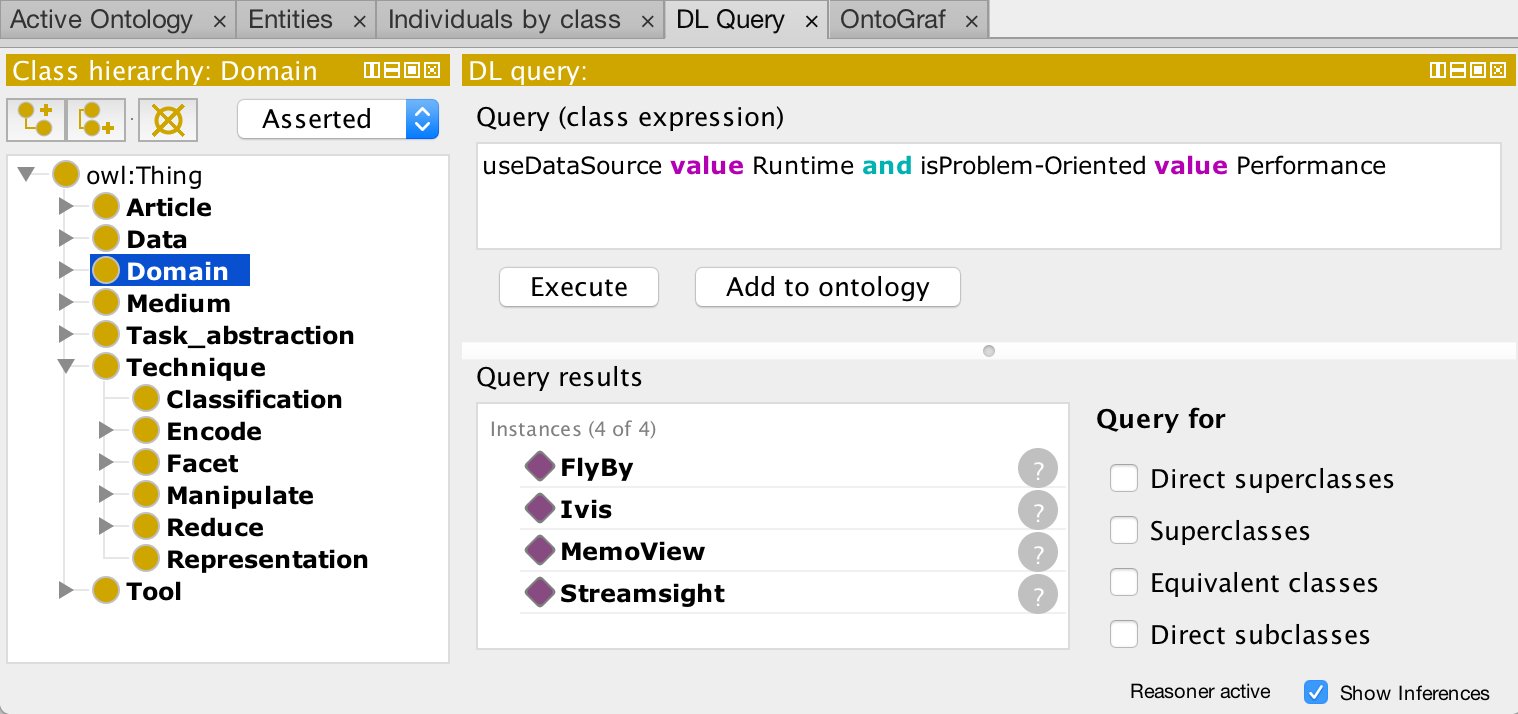}
	\caption{Scenario 1: Finding suitable visualization tools that support the analysis of \emph{performance} issues at \emph{runtime}.}
	\label{fig:protege1}
\end{figure}

Two concepts are defined in the specification of this need: 
\begin{inparaenum}[\itshape (1)\upshape] 
	\item the source of the data is the \emph{runtime} and
	\item the problem dealt is the \emph{performance} of the software system. 
\end{inparaenum}
We translate this specification to the syntax specified by the Ontology Web Language (OWL). Figure~\ref{fig:protege1} shows the resulting query and the suitable tools returned. 

\vspace{0.2cm}
\noindent\emph{\textbf{Scenario 2.} Find visualization tools under a \emph{free} license that support the analysis of source code.}

\begin{figure}[tbp]
	\centering
	\includegraphics[width=\linewidth]{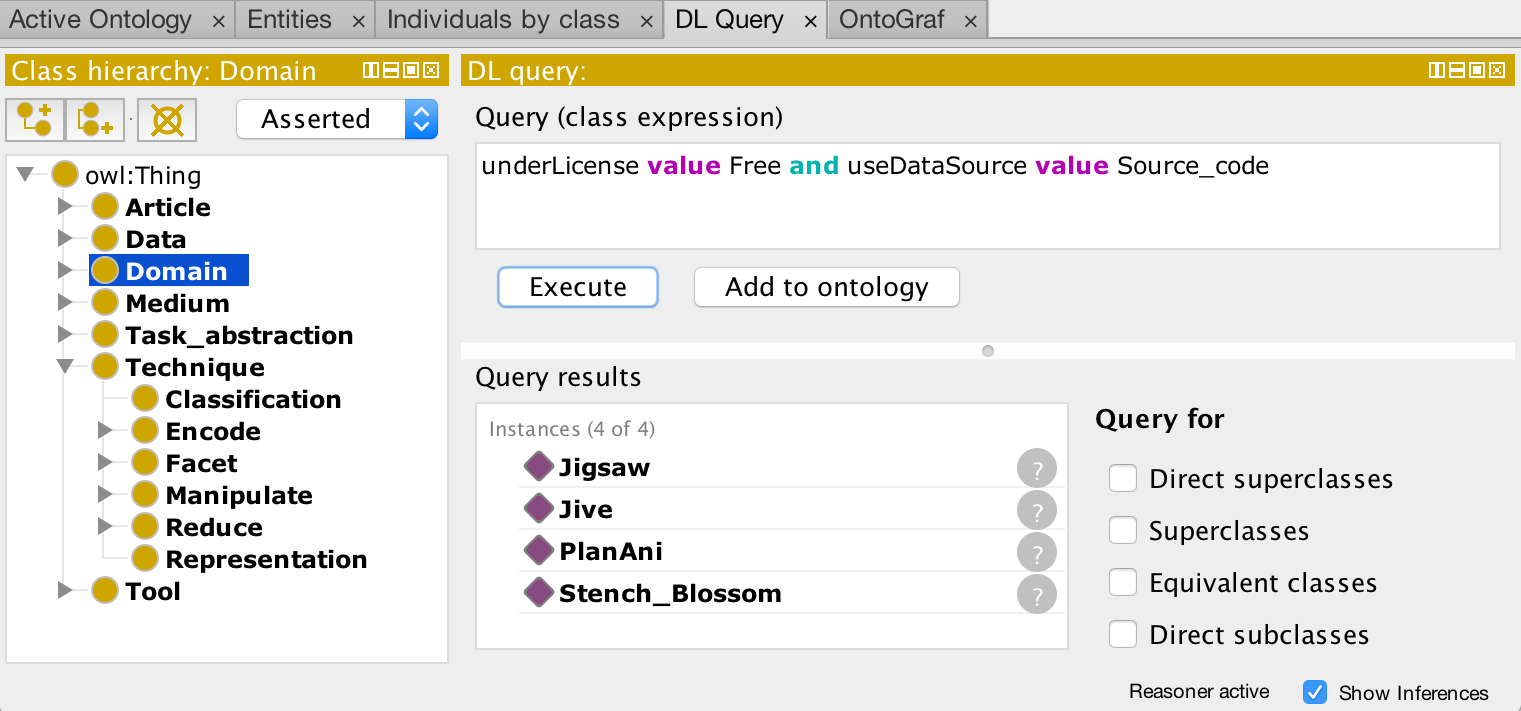}
	\caption{Scenario 2: Finding suitable free visualization tools that support the visualization of source code.}
	\label{fig:protege2}
\end{figure}

Similarly, the specification of this need defines two concepts: 
\begin{inparaenum}[\itshape (1)\upshape] 
	\item the license of the tool has to be \emph{free} and
	\item the source of the data must be the \emph{source code} of the software system. 
\end{inparaenum}
Figure~\ref{fig:protege2} shows the translated specification of the need in the OWL syntax and the suitable tools returned. 

\subsection*{Threats to Validity}
In our paper selection process, we might have overlooked papers from relevant venues that describe important software visualization tools. We mitigated this bias by selecting papers published in the two most frequently cited venues dedicated to software visualization: SOFTVIS and VISSOFT. 
We selected software visualization papers published between 2002 to 2018 in SOFTVIS and VISSOFT.
The excluded papers from other venues or published before 2002 may affect the generalizability of our results. 
We mitigated bias in the data collection procedure (which could obstruct reproducibility of our investigation) by establishing a protocol to extract the data of each paper equally, and by maintaining a spreadsheet to keep records, normalize terms, and identify anomalies.

\section{Conclusion}
\label{sec:conclusion}
Although many software visualization approaches have been proposed to deal with various software concerns, usually developers are not aware of tools they can put into action. In this paper, we presented our attempts to fill the gap between existing software visualizations and their practical applications:
\begin{inparaenum}[\itshape (1)\upshape] 
	\item We presented a curated catalog of 70 available software visualization tools that we linked to their repositories; we classified the tools into various categories (\eg task, data, environment) to help developers who look for suitable visualizations.
	\item We summarized our results in developing VISON, our software visualization ontology. 
\end{inparaenum}

The ontology offers a rich model to encapsulate the various characteristics of software visualizations. We reported on our experience designing and implementing our ontology of software visualizations in the Prot\'eg\'e tool. We demonstrated how the ontology can be used through usage scenarios. Our ontology is publicly available~\cite{Meri19c}. We expect the ontology will help developers find suitable software visualizations and researchers to reflect on the field. Users of the ontology will be able to contribute, for instance, by adding characteristics of new visualizations, or by adding the results of evaluations of existing visualizations. In the future, we plan to combine our previous work on meta-visualization~\cite{Meri16c} with VISON.  

\section*{Acknowledgments}
Merino and Weiskopf acknowledge funding by the Deutsche Forschungsgemeinschaft (DFG, German Research Foundation) -- Project-ID 251654672 -- TRR 161.
Nierstrasz thanks the Swiss National Science Foundation for its financial support of ``Agile Software Assistance'' (project 181973). The authors thank Craig Anslow and Mircea Lungu for valuable comments on a previous version of the paper.

 \bibliographystyle{IEEEtran}
 \bibliography{scg,new}
\end{document}